\documentclass[a4paper,11pt]{article}

\usepackage{amsmath}
\usepackage{amssymb}
\usepackage{amscd}
\usepackage{latexsym}

\def\aa{{\mathcal A}}
\def\bb{{\mathcal B}}
\def\mm{{\mathcal M}}
\def\nn{{\mathcal N}}
\def\dd{{\mathcal D}}
\def\hh{{\mathcal H}}
\def\ll{{\mathcal L}}
\def\ss{{\mathcal S}}

\def\nnn{{\mathbb N}}
\def\rrr{{\mathbb R}}

\def\N{{\mathfrak N}}

\def\T{{\mathfrak T}}

\def\dis{\displaystyle}
\def\llim#1#2{\dis {\lim_{#1\rightarrow #2}}}
\def\Rep{{\rm Rep}}
\def\FRep{{\rm FRep}}
\def\QRep{{\rm QRep}}
\def\FQRep{{\rm FQRep}}
\def\WB{{\tiny {\textsc{w\!b}}}}
\def\Pp{\Pi_p}
\def\pp{\pi_p}

\newcommand{\pa}{partial \mbox{$*$-algebra}}
\newcommand{\tpa}{topological partial \mbox{$*$-algebra}}

\newcommand{\po}{partial O\mbox{$^*$-algebra}}

\newcommand{\ha}{^{*}}
\newcommand{\haa}{^{\rm\textstyle **}}
\newcommand{\ad}{^{\mbox{\scriptsize{\dag}}}}

\def\BR{\relax\ifmmode {\mathbb R} \else${\mathbb R}$\fi}
\def\BN{\relax\ifmmode {\mathbb N} \else${mathbb N} $\fi}
\def\BC{\relax\ifmmode {\mathbb C} \else${mathbb C} $\fi}
\def\B{\relax\ifmmode {\mathcal B}\else${\mathcal B}$\fi}
\def\C{\relax\ifmmode {\mathcal C}\else${\mathcal C}$\fi}
\def\D{\relax\ifmmode {\mathcal D}\else${\mathcal D}$\fi}
\def\E{\relax\ifmmode {\mathcal E}\else${\mathcal E}$\fi}
\def\F{\relax\ifmmode {\mathcal F}\else${\mathcal F}$\fi}
\def\J{\relax\ifmmode {\mathcal J}\else${\mathcal J}$\fi}

\def\H{\relax\ifmmode {\mathcal H}\else${\mathcal H}$\fi}
\def\K{\relax\ifmmode {\mathcal K}\else${\mathcal K}$\fi}
\def\L{\relax\ifmmode {\mathcal L}\else${\mathcal L}$\fi}
\def\N{\relax\ifmmode {\mathcal N}\else${\mathcal N}$\fi}

\def\T{\relax\ifmmode {\mathcal T}\else${\mathcal T}$\fi}
\def\x{\relax\ifmmode {\mbox{*}}\else*\fi}
\def\mm{{\mathcal M}}

\newcommand{\mult}{\;{\scriptscriptstyle \square}\;}
\newcommand{\w}{_{\rm w}}

\newcommand{\LD}{{\L}\w\ad(\D,\H)}
\newcommand{\up}{ \rlap{\raisebox{1.5mm}{$\upharpoonright\,$}} \hskip
2.25pt {\raisebox{-1.0mm}{\vrule height 10 pt width 0.5 pt}\ }}%

\begin{document}
\begin{center}
{\Large
Unbounded C$^*$-seminorms and $*$-Representations of Partial $*$-Algebras}

{\large
F. Bagarello, A. Inoue and C. Trapani
}
\end{center}

\begin{abstract}

The main purpose of this paper is to construct $*$-representations from
unbounded C$^*$-seminorms on partial $*$-algebras and to investigate their
$*$-representations.
\end{abstract}
\newpage
\setcounter{section}{0}
\section{Introduction and Preliminaries}
A C*-seminorm $p$ on a locally convex *-algebra $\mathcal A$ is a seminorm
enjoying the C*-property $p(x^*x)= p(x)^2$, $x \in \mathcal A$. They have been
extensively studied in the literature; see, e.g. [9-13, 19]. One of the main
points
of the theory is that every *-representation of the completion $(\mathcal A,p)
$ is bounded.

Generalizations of this notions have led Bhatt, Ogi and one of us \cite{bio3}
to consider the so-called {\em unbounded C*-seminorms} on *-algebras. Their
main feature is that they need not be defined on the whole $\mathcal A$ but
only on a *-subalgebra of it. This fact allows the existence of unbounded
representations of $\mathcal A$ (and motivates the adjective "unbounded" used
to name them). But it is not only for need of mathematical generalization that
it makes sense to consider unbounded C*-seminorms but also because of it
appearance  in some subject of mathematical physics [1,15,18].
However, when considering unbounded C*-seminorms on a locally convex *-algebra
$\mathcal A$ whose multiplication is not jointly continuous one is naturally
led to consider partial algebraic structures: in that case in fact the
completion of $\mathcal A$ is no longer, in general, a locally convex
*-algebra but only  a topological quasi *-algebra \cite{lass1, lass2}. Quasi
*-algebras are a particular case of partial *-algebras \cite{ak} . Roughly
speaking, a partial *-algebra $\mathcal A$ is a linear space with involution
and a partial multiplication defined on a subset $\Gamma$ of $\mathcal A
\times \mathcal A$ enjoying some of the usual properties of the
multiplication, with the very relevant exception of associativity. Of course,
as one of the main tools in the study of *-algebras is the theory of
*-representations, partial *-algebras of operators (the so called partial
O*-algebras) have been considered as the main instance of these {\em new}
algebraic structures and a systematic study has been undertaken [3-6]. From
a more abstract point of view, the possibility of introducing topologies
compatible with the structure of a partial *-algebra has been investigated in
\cite{abt}.

The present paper is organized as follows. In Section 2, starting from a
C*-seminorm $p$ on a partial *-algebra $\mathcal A$ we prove the existence of
{\em quasi *-representations} of $\mathcal A$; they are named in this way
since the usual rule for the multiplication, holds in a sense that remind the
multiplication in quasi *-algebras.
These quasi *-representations depend essentially on a certain subspace
${\mathcal N}_p$ of the domain ${\mathcal D}(p)$ of the C*-seminorm $p$. Of
course, by adding assumptions on  ${\mathcal N}_p$ we are led to consider a
variety of situations of some interest. In this perspective, we introduce the
notions of finite and (weakly-) semifinite C*-seminorms and study in detail
the quasi *-representations that they induce.

In Section 3 we consider the problem as to whether a *-representation of
$\mathcal A$,  in the sense of \cite{ait2}, does really exist  or in other
words if the quasi *-representation, whose existence has been proved in
section 2, is indeed a *-representation.

In Section 4, we reverse the point of view: starting from a *-representation
$\pi$ of a partial *-algebra, we construct an unbounded C*-seminorm $r_\pi$ on
$\mathcal A$ which turns out to admit a *-representation $\pi^{N}_{r_\pi}$
called {\em natural}. We then investigate the relationship between
$\pi^{N}_{r_\pi}$ and the *-representation $\pi$ where we had started with.

Section 5 is devoted to the discussion of some examples.

\vspace{2mm}
Before going forth, we shortly give some definitions needed in the sequel.

A {\em \pa} is a complex vector
space $\aa$, endowed with an involution $x \mapsto x\ha$ (that is, a
bijection such that $x\haa = x$)
and a partial multiplication defined by a set $\Gamma \subset \aa \times
\aa$ (a binary relation) such that:

$\;$(i) $(x,y) \in \Gamma$ implies $(y\ha,x\ha) \in \Gamma$;

$\,$(ii) $(x,y_1), (x,y_2) \in \Gamma$ implies $(x, \lambda y_1 + \mu y_2)
\in \Gamma, \, \forall \,\lambda,\mu \in \BC; $

(iii) for any $(x,y) \in \Gamma$, there is defined a product $x \, y \in
\aa$, which is distributive with respect to the addition and satisfies the
relation $ (x\,y)\ha =
y\ha \, x\ha$.

\noindent
 The element $e$ of the \pa\ $\aa$ is called a unit if
$e\ha = e, \, (e,x) \in \Gamma, \, \forall \, x \in \aa$, and $ e \, x = x
\, e = x, \, \forall \, x \in \aa$.

Given the defining set $\Gamma$, spaces of multipliers are defined in the
obvious way:
\begin{center}
\begin{tabular}[t]{lll}
$(x,y) \in \Gamma$ & $\Leftrightarrow$ & $x \in L(y) \, \mbox{ or } x
\mbox{ is a left multiplier of } y $ \\
& $\Leftrightarrow$ & $y\in R(x) \, \mbox{ or } y \mbox{ is a right
multiplier of } x $.
\end{tabular}
\end{center}

Notice that the partial multiplication is {\em not} required to be
associative (and
often it is not). The following weaker notion is therefore in use:
a \pa\ $\aa$ is said to be {\em semi-associative} if $y \in R(x)$ implies $y
\cdot z \in R(x)$ for every $z \in R(\aa) $ and
\begin{equation}
(x \cdot y) \cdot z = x \cdot (y \cdot z).
\end{equation}

Let $\aa [\tau]$ be a \pa, which is a topological vector space for the locally
convex topology $\tau$.  Then
$\aa [\tau]$ is called a {\em  \tpa} if the following two conditions are
satisfied \cite{abt}:

\begin{itemize}
\item[(i)]
 the involution $a \mapsto a\x$ is $\tau$-continuous;

\item[(ii)]
 the maps $a \mapsto xa$ and  $a \mapsto ay$ are $\tau$-continuous for all $x
\in L(\aa)$ and $y \in R(\aa)$.
\end{itemize}

A {\em quasi *-algebra} $(\aa, \aa_0)$ is a \pa\ where the multiplication is
defined via the *-algebra $\aa_0 \subset \aa$ by taking $\Gamma$ as
$$ \Gamma = \{(a,b)\in \aa \times \aa: a \in \aa_0 \mbox{ or } b \in \aa_0\} .
$$
If $\aa$ is endowed with a locally convex topology which makes it into a
topological partial *-algebra and $\aa_0$ is dense in $\aa$, then $(\aa,
\aa_0)$ is said to be a {\em topological} quasi *-algebra.

We turn now to \po s.
Let $\H$ be a complex Hilbert space and $\D$ a dense subspace of $\H$. We
denote by $ \L\ad(\D,\H) $ the set of all (closable) linear operators $X$
such that $ {D}(X) = {\D},\; {D}(X\x) \supseteq {\D}.$ The set $
\L\ad(\D,\H ) $ is a \pa\
with respect to the following operations : the usual sum $X_1 + X_2 $, the
scalar multiplication $\lambda X$,
the involution $ X \mapsto X\ad = X\x \up {\D}$ and the {\em (weak)}
partial multiplication
$X_1 \mult X_2 = {X_1}\ad\x X_2$, defined whenever $X_2$ is a weak right
multiplier of $X_1$ (equivalently, $X_1$ is a weak left multiplier of
$X_2$), that is, iff $ X_2 {\D}\subset {D}({X_1}\ad\x)$ and $ X_1\x {\D}
\subset {D}(X_2\x)$
(we write $ X_2 \in R^{\rm w}(X_1)\; \mbox{or} \; X_1 \in L^{\rm w}(X_2)$).
When we regard $ \L\ad(\D,\H) $ as a \pa\ with those operations, we denote
it by $\LD $.

A \po\ on \D\ is a *-subalgebra $\mm$ of $\LD $,
that is, $\mm$ is a subspace of $\LD $, containing the identity and such
that $X\ad \in \mm $ whenever $X \in \mm $ and $X_1 \mult X_2 \in \mm$ for
any $X_1, X_2 \in \mm$ such that $X_2 \in R^{\rm w} (X_1).$ Thus $\LD$
itself is the largest \po\ on the domain \D.

Given a $\ad$-invariant subset $\nn$ of $\L\ad(\D,\H)$,  the familiar weak
bounded commutant, is defined as follows: $$
\nn'\w = \{ C \in {\mathcal B}(\H) ; (CX \xi| \eta) = (C \xi|X\ad \eta) \;
\mbox{for each} \; \xi,\eta \in \D \; \mbox{and} \; X \in \nn \}. $$
The last definitions we need are related with representations.

A \emph{*-representation} of a \pa\ $\aa$ is a *-homomorphism of $\aa$ into
${\L}\w\ad(\D(\pi),\H_{\pi})$, for some pair $\D(\pi) \subset \H_{\pi}$, that
is, a linear map $\pi : \aa \to
{\L}\w\ad(\D(\pi),\H_{\pi})$ such that : (i) $\pi(x\ha) = \pi(x)\ad$ for every
$x \in \aa$; (ii) $x
\in L(y)$ in $\aa$ implies $\pi(x) \in L^{\rm w}(\pi(y))$ and $\pi(x) \mult
\pi(y)
= \pi(xy)$.
If $\pi$ is a *-representation of the \pa\ $\aa$ into
${\L}\w\ad(\D(\pi),\H_{\pi})$, we define $ \widetilde{\D(\pi)}$ as the
completion of $\D(\pi)$ with respect to the graph topology defined by
$\pi(\aa)$.
Furthermore we put:
$$ \widehat{\D(\pi)}= \bigcap_{x \in \aa}\D(\overline{\pi(x)})$$
$$ \D(\pi)^\ast =\bigcap_{x \in \aa}\D(\pi(x)^\ast).$$
We say that $\pi$ is: {\em closed} if $\D(\pi)=\widetilde{\D(\pi)}$; {\em
fully-closed} if $\D(\pi)= \widehat{\D(\pi)}$; {\em self-adjoint} if
$\D(\pi)=\D(\pi)^\ast$.\\
Let $\pi_1$ and $\pi_2$ be *-representations of $\aa$. With the notation
$\pi_1 \subset \pi_2$ we mean that $\H_{\pi_1} \subseteq \H_{\pi_2}$;
$\D({\pi_1}) \subseteq \D({\pi_2})$ and $\pi_1(a)\xi=\pi_2(a)\xi$ for each
$\xi \in \D({\pi_1})$.

By considering the identical *-representations, the terms {\em fully-closed,
self-adjoint}, etc., can also be referred to a \po\ on a given domain $\D$ and
then generalized, in obvious way, to an arbitrary $\ad$-invariant subset of
$\L\ad(\D,\H)$.

\section{ Representations induced by unbounded \\
C$^*$-seminorms}
In this section we construct (quasi) $*$-representations of partial
$*$-algebras from unbounded C$^*$-seminorms.
Throughout this paper we treat only with partial $*$-algebras whose partial
multiplication satisfies the properties:
\begin{equation*}
\text{(A)} \hspace{12mm}
\begin{cases}
y^*(ax) = (y^* a)x, \\
a(xy) = (ax) y, \forall a \in \aa, \forall x, y \in R(\aa).
\end{cases}
\end{equation*}
We remark that if $\aa$ is semi-associative then it satisfies Property (A).

\

{\bf Definition 2.1.}
A mapping $p$ of a (partial) $*$-subalgebra $\dd(p)$ of $\aa$ into $\rrr^+$ is
said to be an {\it unbounded} $m^*$-(semi)norm on $\aa$ if

(i) $p$ is a (semi) norm on $\dd(p)$;

(ii) $p(x^*)=p(x), \ \ \forall x \in \dd(p)$;

(iii) $ p(xy) \leq p(x) p(y), \ \ \forall x, y \in \dd(p) \text{ s.t. } x \in
L(y)$.
\\
An unbounded $m^*$-(semi)norm $p$ on $\aa$ is said to be an unbounded
C$^*$-(semi)norm if

(iv) $p(x^*x) = p(x)^2, \ \ \forall x \in \dd(p) \text{ s.t. } x^* \in L(x)$.
\\
An unbounded $m^*$-(semi)norm (resp. C$^*$-(semi)norm) on $\aa$ is said to be
a $m^*${\it -(semi)norm } (resp. {\it C$^*$-(semi)norm}) if $\dd(p) = \aa$.
\\
An (unbounded) $m^*$-seminorm $p$ on $\aa$ is said to have Property (B) if it
satisfies the following basic density-condition:

(B) \hspace*{6mm}
$R(\aa) \cap \dd(p)$ is total in $\dd(p)$ with respect to $p$.

\

{\bf Lemma 2.2.}
Let $p$ be a $m^*$-seminorm on $\aa$ having Property (B), that is, $R(\aa)$ is
$p$-dense in $\aa$. We denote by $\hat{\aa}$ the set of all Cauchy sequences
in $\aa$ w.r.t. the seminorm $p$ and define an equivalent relation in
$\hat{\aa}$ as follows:\\
$\{ a_n \} \sim \{b_n \}$ iff $\llim{n}{\infty} p(a_n - b_n) = 0$. Then the
following statements hold:

(1) The quotient space $\hat{\aa} /\sim$ is a Banach $*$-algebra under the
following operations, involution and norm:
\begin{align*}
&\{ a_n \}^\sim + \{ b_n \}^\sim \equiv \{ a_n + b_n \}^\sim ; \\
&\lambda \{ a_n \}^\sim \equiv \{ \lambda a_n \}^\sim ;\\
&\{ a_n\}^\sim \{ b_n \}^\sim \equiv \{ x_n y_n \}^\sim, \text{ where } \{ x_n
\}^\sim \text{ and } \{ y_n \}^\sim \text{ in } R(\aa) \text{ s.t. } \\
&\hspace*{45mm} \{ x_n \}^\sim \equiv \{ a_n \}^\sim \text{ and } \{ y_n
\}^\sim \equiv \{ b_n \}^\sim;\\
&\{ a_n \}^{\sim *} \equiv \{ a_n^* \}^\sim \\
&\| \{ a_n \}^\sim \|_p \equiv = \llim{n}{\infty} p(a_n).
\end{align*}
(2)  For each $a \in \aa$ we put
\begin{align*}
& \tilde{a} = \{ a_n \}^\sim \hspace{6mm} (a_n = a, n \in \nnn), \\
& \tilde{\aa} = \{ \tilde{a} ; a \in \aa \}.
\end{align*}
Then $\tilde{\aa} $ is a dense $*$-invariant subspace of $\hat{\aa} /\sim $
satisfying $\tilde{\mathstrut a} \tilde{\mathstrut b} = (ab)^\sim$ whenever $a
\in L(b)$.
\\
(3) Suppose $p$ is a C$^*$-seminorm on $\aa$. Then $\hat{\aa}/\sim $ is a
C$^*$-algebra.\\[3mm]
\indent
{\bf Proof.} As in the usual construction of the completion of a normed space,
it can be shown that $\hat{\aa}/\sim$ is a Banach space.\\
(1) We first show that $\{ a_n \}^\sim \{ b_n \}^\sim$ is well-defined and the
relation defines a multiplication of $\hat{\aa}/\sim$.
Since $R(\aa)$ is $p$-dense in $\aa$, for each $\{ a_n \}, \{ b_n \} \in
\hat{\aa}$ there exist sequences $\{ x_n \}, \{ y_n \}$ in $R(\aa)$ such that
$\{ a_n \}^\sim = \{ x_n \}^\sim$ and $\{ b_n \}^\sim = \{ y_n \}^\sim$.
Then it follows from the submultiplicativity of $p$ that $\{ x_n y_n \}^\sim
\in \hat{\aa}$ and $\{ a_n \}^\sim \{ b_n \}^\sim$ is independent of the
choice of the equivalent sequences $\{ x_n \}$ and $\{ y_n \}$.
Further, the relation $\{ a_n \}^\sim \{ b_n \}^\sim$ defines a multiplication
of $\hat{\aa}/\sim$. In fact, the associativity follows from the equations:
\begin{align*}
\{ a_n \}^\sim ( \{ b_n \}^\sim \{ c_n \}^\sim)
&= \{ x_n \}^\sim ( \{ y_n z_n \}^\sim) \\
&= \{ x_n (y_n z_n) \}^\sim \\
&= \{ (x_n y_n )z_n \}^\sim \\
&= (\{ a_n \}^\sim \{ b_n \}^\sim ) \{ c_n \}^\sim,
\end{align*}
where $\{ x_n \}, \{ y_n \}, \{ z_n \} \subset R(\aa) $ s.t. $\{ x_n \}^\sim =
\{ a_n \}^\sim, \{ y_n \}^\sim = \{ b_n \}^\sim$ and $\{ z_n \}^\sim = \{ c_n
\}^\sim$, and the others can be proved in similar way.
Thus $\hat{\aa}/\sim$ is a usual algebra. We next show that $\{ a_n \}^\sim
\mapsto \{ a_n^* \}^\sim$ is an involution of the algebra $\hat{\aa}/\sim$.
Take arbitrary $\{ a_n \}, \{ b_n \} \in \hat{\aa}$.
Since $R(\aa)$ is $p$-dense in $\aa$ and $p(a^*)= p(a), \forall a \in \aa$,
there exist sequences $\{ x_n \}, \{x_n' \}, \{ y_n \}, \{ y_n' \}$ in
$R(\aa)$ such that $\{ a_n \}^\sim = \{ x_n \}^\sim, \{ a_n^* \}^\sim = \{
x_n^* \}^\sim = \{ x_n' \}^\sim, \{ b_n \}^\sim= \{ y_n \}^\sim$ and
$\{ b_n^* \}^\sim = \{ y_n^* \}^\sim = \{ y_n' \}^\sim $. Then we have
\begin{align*}
( \{ a_n \}^\sim \{ b_n \}^\sim )^* = \{ x_n y_n \}^{\sim *}
&= \{ y_n^* x_n^* \}^\sim \\
&= \{ y_n' x_n' \}^\sim \\
&= \{ b_n^* \}^\sim \{ a_n^* \}^\sim \\
&= \{ b_n \}^{\sim *} \{ a_n \}^{\sim *}.
\end{align*}
The others can be proved in similar way.
Thus $\hat{\aa}/\sim$ is a $*$-algebra.
Further, we have
\begin{align*}
\| \{ a_n \}^\sim \{ b_n \}^\sim \|_p = \| \{ x_n y_n \} \|_p
&= \llim{n}{\infty} p(x_n y_n) \\
&\leq \llim{n}{\infty} p(x_n) p(y_n) \\
&= \| \{ a_n \}^\sim \|_p \| \{ b_n \}^\sim \|_p,
\end{align*}
\begin{align*}
\| \{ a_n \}^{\sim *} \|_p = \| \{ a_n^* \}^\sim \|_p
&= \llim{n}{\infty} p(a_n^*) \\
&= \llim{n}{\infty} p(a_n ) \\
&= \| \{ a_n \}^\sim \|_p
\end{align*}
for each $\{ a_n \}, \{ b_n \} \in \hat{\aa}$, which implies that
$\hat{\aa}/\sim$ is a Banach $*$-algebra.
The statements (2) and (3) can be proved in similar way. This completes the
proof.

\

>From now on we denote as $\Sigma (\aa)$ the set of all unbounded C*-seminorms
on $\aa$ and with $\Sigma_B(\aa)$ the subset of $\Sigma (\aa)$ consisting of
those satisfying Property (B).

Let $p$ be an unbounded C$^*$-seminorm on $\aa$ having Property (B), i.e. $p
\in \Sigma_B(\aa)$. By Lemma 2.2, $\aa_p \equiv \widehat{\dd(p)} /\sim$ is a
C$^*$-algebra. We denote by $\Rep(\aa_p)$ the set of all $*$-representations
$\Pi_p$ of the C$^*$-algebra $\aa_p$ on Hilbert space $\hh_{\Pi_p}$, and put
\[
\FRep (\aa_p) = \{ \Pi_p \in \Rep (\aa_p) ; \Pi_p \text{ is faithful} \}.
\]
Then we have the following

\

{\bf Proposition 2.3.} For any $\Pi_p \in \Rep(\aa_p)$ we put
\[
\pi_p^\circ (x) = \Pi_p( \tilde{x} ), \ \ \ x \in \dd(p).
\]
Then $\pi_p^\circ $ is a $*$-representation of $\dd(p)$ on $\hh_{\Pi_p}$.

\

The previous proposition provides the most natural way to define a
*-representation of $\dd(p)$. However $\pi_p^\circ$ cannot be extended to the
whole $\aa$. The construction of *-representations of $\aa$ requires a more
detailed analysis. This will be the content of the next propositions.
To begin with, we put
\[
\N_p = \{ x \in \dd(p) \cap R(\aa); ax \in \dd(p), \forall a \in \aa \}.
\]
Then we have the following

\

{\bf Lemma 2.4.} (1) $\N_p$ is an algebra satisfying $(\dd(p) \cap R(\aa))
\N_p \subset \N_p$.
\\
(2) We denote by $\T_p$ the closure of $\tilde{\N_p}$ in the C$^*$-algebra
$\aa_p$. Then $\T_p$ is a closed left ideal of $\aa_p$.\\
(3) $\Pi_p ({\tilde{\N_p}}^2 ) \hh_{\Pi_p}$ is dense in $\Pi_p(\tilde{\N_p})
\hh_{\Pi_p}$. \\[3mm]
\indent
{\bf Proof.}
(1) This follows from the semi-associativity (A). \\
(2) Since $\dd(p) \cap R(\aa)$ is $p$-dense in $\dd(p)$ and the above (1), it
follows that $\dd(p)^\sim \N_p^\sim \subset \T_p$, and so $\dd(p)^\sim \T_p
\subset \T_p$. Since $\dd(p)^\sim$ is dense in the C$^*$-algebra $\aa_p$, we
have $\aa_p \T_p \subset \T_p$.\\
(3) It is clear that $\Pi_p({\tilde{\N_p}}^2) \hh_{\Pi_p}$ is dense in
$\Pi_p(\tilde{\N_p} \T_p) \hh_{\Pi_p}$.
Since $\T_p$ is a closed left ideal of the C$^*$-algebra $\aa_p$, there exists
a direct net $\{ U_\lambda \}$ in $\T_p$ such that $\dis \lim_\lambda \| A
U_\lambda - A \|_p = 0$ for each $A \in \T_p$, which implies that
$\Pi_p(\tilde{\N_p} \T_p) \hh_{\Pi_p}$ is dense in $\Pi_p(\tilde{\N_p})
\hh_{\Pi_p}$. Hence $\Pi_p( {\tilde{\N_p}}^2) \hh_{\Pi_p}$ is dense in
$\Pi_p(\tilde{\N_p}) \hh_{\Pi_p}$.

\

Let now
$ \dd(\pi_p)$  be the linear span of $ \{ \Pi_p ((xy)^\sim) \xi ; x, y \in
\N_p, \xi \in \hh_{\Pi_p} \}$
and $\hh_{\pi_p}$ be the closure of $\dd(\pi_p)$ in $\hh_{\Pi_p}$; we define
\begin{align*}
& \pi_p(a) \left( \sum\limits_{k} \Pi_p( (x_k y_k)^\sim) \xi_k \right)
= \sum\limits_{k} \Pi_p ( (a x_k)^\sim \widetilde{y_k} ) \xi_k, \\
& \hspace{45mm} a \in \aa, \sum\limits_{k} \Pi_p ((x_k y_k)^\sim) \xi_k \in
\dd(\pi_p).
\end{align*}
{\bf Remark.}  By Lemma 2.4, (3) we have
\begin{align*}
\hh_{\pi_p}
&\equiv \text{ closed linear span of } \{ \Pi_p(\tilde{x_1} \tilde{x_2}) \xi ;
x_1, x_2 \in \nn_p, \xi \in \hh_{\Pi_p} \} \\
&= \text{ closed linear span of } \{ \Pi_p(\tilde{x}) \xi ; x \in \nn_p, \xi
\in \hh_{\Pi_p} \}.
\end{align*}
In general, it may happen that $\hh_{\pi_p}$ is very 'small' compared to
$\hh_{\Pi_p}$. This point will be considered at the end of this Section, where
{\em well-behaved} representations related to unbounded C*-seminorms will be
introduced.

\
Now, we prove the following

\

{\bf Lemma 2.5.}
$\pi_p$ is a linear map of $\aa$ into $\ll^\dagger (\dd(\pi_p), \hh_{\pi_p})$
satisfying the following properties:

(i) $\pi_p(a^*) = \pi_p(a)^\dagger, \hspace{5mm} \forall a \in \aa$;

(ii) $\pi_p(ax) = \pi_p(a)${\tiny $\Box$}$\pi_p(x),
\hspace{5mm} \forall a \in \aa, \forall x \in R(\aa)
$;

(iii) $\| \overline{\pi_p(x)} \| \leq p(x), \hspace{5mm} \forall x \in
\dd(p)$.
Further, if $\pi_p \in \FRep (\aa_p)$ then
$ \| \overline{\pi_p(x)} \| = p(x), \hspace{5mm} \forall x \in \N_p.$\\[3mm]
\indent
{\bf Proof.}
By Lemma 2.4, (2), (3) we have
\[
\Pi_p((ax)^\sim \tilde{y}) \xi \in \Pi_p (\T_p) \hh_{\Pi_p} \subset
\hh_{\pi_p}
\]
for each $a \in \aa$, $x, y \in \N_p$ and $\xi \in \hh_{\Pi_p}$, and further
by Property (A)
\begin{align*}
(\Pi_p((ax_1)^\sim \tilde{y_1} ) \xi | \Pi_p ( \tilde{x_2} \tilde{y_2} ) \eta)
& = (\Pi_p(\tilde{y_1}) \xi | \Pi_p ( ((ax_1)^* x_2)^\sim) \Pi_p( \tilde{y_2}
) \eta)
\\
&= (\Pi_p (\tilde{y_1}) \xi | \Pi_p ((x_1^* (a^* x_2))^\sim ) \Pi_p
(\tilde{y_2} ) \eta) \\
&= (\Pi_p (\tilde{x_1} \tilde{y_1} ) \xi | \Pi_p ((a^* x_2)^\sim \tilde{y_2})
\eta)
\end{align*}
for each $a \in \aa$, $x_1, y_1, x_2, y_2 \in \N_p$ and $\xi, \eta \in
\hh_{\Pi_p}$, which implies that $\pi_p(a)$ is a well-defined linear map from
$\dd(\pi_p)$ to $\hh_{\pi_p}$ satisfying $\pi_p(a^*) = \pi_p(a)^\dagger$. It
is clear that $\pi_p$ is a linear map of $\aa$ into $\ll^\dagger(\dd(\pi_p),
\hh_{\pi_p})$.
We next show the statement (ii).
Take arbitrary $a \in \aa$ and $x \in R(\aa)$.
By Property (A) we have
\[
z^* ((ax)y) = (z^* (ax)) y = ((z^* a) x) y
\]
for each $a \in \aa, x \in R(\aa)$ and $y, z \in \N_p$, and hence it follows
from Lemma 2.2, (2) that
\begin{align*}
(\pi_p(ax) \Pi_p( \tilde{y_1} \tilde{y_2} ) \xi | \Pi_p (\tilde{z_1}
\tilde{z_2} ) \eta)
&= (\Pi_p((ax) y_1)^\sim ) \Pi_p(\tilde{y_2}) \xi | \Pi_p(\tilde{z_1} )
\Pi_p(\tilde{z_2} ) \eta ) \\
&= (\Pi_p(( (z_1^* a) x)^\sim \tilde{y_1}) \Pi_p(\tilde{y_2}) \xi |
\Pi_p(\tilde{z_2}) \eta) \\
&= (\Pi_p (\tilde{y_1} \tilde{y_2} ) \xi | \Pi_p (\tilde{x^*} (z_1^* a)^{*
\sim}) \Pi_p (\tilde{z_2}) \eta) \\
&= (\Pi_p (x) \Pi_p (\tilde{y_1} \tilde{y_2}) \xi | \pi_p(a)^\dagger \Pi_p
(\tilde{z_1} \tilde{z_2}) \eta)\\
&= (\pi_p (x) \Pi_p (\tilde{y_1} \tilde{y_2}) \xi | \pi_p(a)^\dagger \Pi_p
(\tilde{z_1} \tilde{z_2}) \eta)
\end{align*}
for each $y_1, y_2, z_1, z_2 \in \N_p$ and $\xi, \eta \in \hh_{\Pi_p}$, which
implies the statement (ii).
Take an arbitrary $x \in \dd(p)$.
Since $\overline{\pi_p(x)} = \Pi_p(\tilde{x}) \up \hh_{\pi_p}$, it follows
that
$\| \overline{\pi_p(x)} \| \leq \| \Pi_p (\tilde{x}) \| = p(x)$.
Suppose $\Pi_p \in \FRep (\aa_p)$.
Take an arbitrary $x \in \N_p$. It is sufficient to show $\|
\overline{\pi_p(x)} \| \geq p(x)$. If $p(x) = 0$, then this is obvious.
Suppose $p(x) \neq 0$.
We put $y = x /p(x)$. Since
\[
\| \Pi_p(\tilde{y}) \xi \| \leq \| \Pi_p(\tilde{y}) \| \| \xi \| = p(y) \| \xi
\| \leq 1
\]
for each $\xi \in \hh_{\Pi_p} $ s.t. $\| \xi \| \leq 1 $ and
$\Pi_p(\tilde{\N_p}) \hh_{\Pi_p}$ is total in $\hh_{\pi_p}$ (by Lemma 2.4,
(3)and the Remark thereafter), it follows that
\begin{align*}
\| \overline{\pi_p (y)} \| = \| \overline{\pi_p(y^*)} \|
&\geq \sup \{ \| \pi_p(y^*) \Pi_p(\tilde{y}) \xi \|; \xi \in \hh_{\Pi_p}
\text{ s.t. } \| \xi \| \leq 1 \} \\
&= \sup \{ \| \Pi_p( (y^* y)^\sim ) \xi \| ; \xi \in \hh_{\Pi_p} \text{ s.t. }
\| \xi \| \leq 1 \} \\
&= \| \Pi_p ((y^* y)^\sim) \| \\
&= p(y^* y) = p(y)^2 = 1,
\end{align*}
which implies that $\| \overline{\pi_p(x)} \| \geq p(x)$.
This completes the proof.

\

{\bf Remark.} If, instead of following the above procedure, we would have
taken
\begin{align*}
& \dd(\pi) = \text{ linear span of } \{ \Pi_p (\tilde{x}) \xi ; x \in
\N_p, \xi \in \hh_{\Pi_p} \}, \\
& \hh_{\pi} = \text{ the closure of } \dd(\pi) \text{ in }
\hh_{\Pi_p}, \\
& \pi(a) \left( \sum_k \Pi_p(\tilde{x_k} ) \xi_k \right) = \sum_k
\Pi_p((ax_k)^\sim) \xi_k, \\
& \hspace*{50mm} a \in \aa, \sum_k \Pi_p(\tilde{x_k}) \xi_k \in
\dd(\pi),
\end{align*}
then we could not conclude that $\pi(a)$ belongs to $\ll^\dagger
(\dd(\pi), \hh_{\pi})$ for each $a \in \aa$.

\

So far, we don't know whether $\pi_p$ is a $*$-representation of $\aa$ for the
lack of semi-associativity of partial multiplication, and so we define the
following notion:

\

{\bf Definition 2.6.} A linear map $\pi$ of $\aa$ into $\ll^\dagger (\dd(\pi),
\hh_{\pi})$ is said to be a {\it quasi $*$-representation} if

(i) $\pi(a^*) = \pi(a)^\dagger, \forall a \in \aa$,

(ii) $ \pi(ax) = \pi(a)${\tiny $\Box$}$\pi(x), \forall a \in \aa, \forall x
\in R(\aa)$.

\ \\
By Lemma 2.5, for each $p \in \Sigma_B(\aa)$, every $\pi_p$ is a quasi
$*$-representation of $\aa$, and it is said to be a {\it quasi
$*$-representation of $\aa$ induced by $p$}.

\

We summerize in the following scheme the method of construction $\pp$ from an
unbounded C$^*$-seminorm $p$ desctibed above:

\vskip-5mm
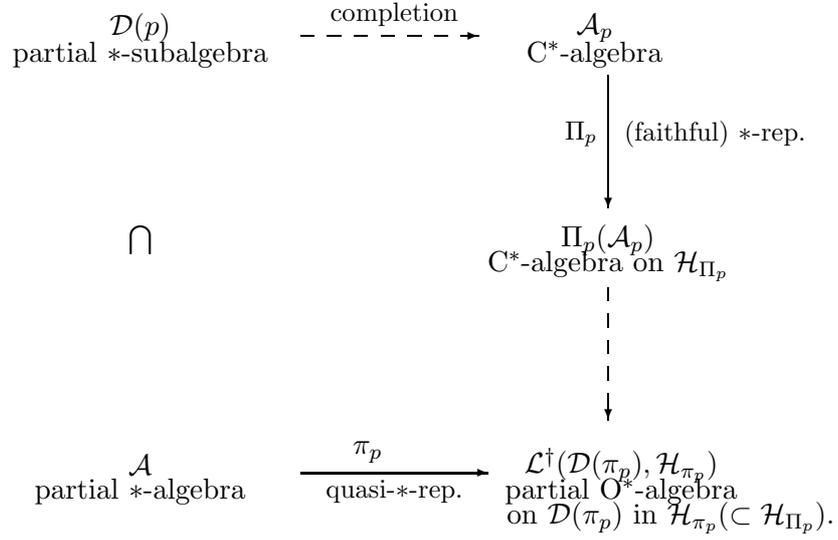
\begin{picture}(95,40)
\put(90,0){\makebox(10,6)[c]{$\dd(p)$}}
\put(90,-10){\makebox(10,6)[c]{partial $*$-subalgebra}}
\put(185,5){\makebox(10,6)[c]{\small completion}}
\put(155,0){\line(1,0){5}}
\put(165,0){\line(1,0){5}}
\put(175,0){\line(1,0){5}}
\put(185,0){\line(1,0){5}}
\put(195,0){\line(1,0){5}}
\put(205,0){\line(1,0){5}}
\put(215,0){\line(1,0){5}}
\put(220,0){\vector(1,0){1.5}}
\put(260,0){\makebox(10,6)[c]{$\aa_p$}}
\put(260,-10){\makebox(10,6)[c]{C$^*$-algebra}}
\put(270,-15){\vector(0,-1){50}}
\put(255,-40){\makebox(10,6)[c]{{\small $\Pp$}}}
\put(305,-40){\makebox(10,6)[c]{{\small (faithful) $*$-rep.}}}
\put(265,-80){\makebox(10,6)[c]{$\Pp(\aa_p)$}}
\put(265,-90){\makebox(10,6)[c]{C$^*$-algebra on $\hh_{\Pi_p}$}}
\put(90,-80){\makebox(10,6)[c]{$\bigcap$}}
\put(270,-95){\line(0,-1){5}}
\put(270,-105){\line(0,-1){5}}
\put(270,-115){\line(0,-1){5}}
\put(270,-125){\line(0,-1){5}}
\put(270,-135){\line(0,-1){5}}
\put(270,-140){\vector(0,-1){5}}
\put(270,-165){\makebox(10,6)[c]{$\ll^\dagger(\dd(\pp), \hh_{\pp})$}}
\put(270,-175){\makebox(10,6)[c]{partial O$^*$-algebra}}
\put(270,-185){\makebox(10,6)[c]{\hspace*{13mm}on $\dd(\pp)$ in
$\hh_{\pp}(\subset \hh_{\Pp})$.}}
\put(90,-165){\makebox(10,6)[c]{$\aa$}}
\put(90,-175){\makebox(10,6)[c]{partial $*$-algebra}}
\put(155,-165){\vector(1,0){70}}
\put(175,-160){\makebox(10,6)[c]{$\pp$}}
\put(185,-175){\makebox(10,6)[c]{{\small quasi-$*$-rep.}}}
\end{picture}

\vskip70mm

Here the arrow $A$
\begin{picture}(26,20)
\put(0,3){\line(1,0){3}}
\put(6,3){\line(1,0){3}}
\put(12,3){\line(1,0){3}}
\put(18,3){\line(1,0){3}}
\put(21,3){\vector(1,0){5}}
\end{picture}
$B$ means that $B$ is constructed from $A$.

\

We put
\begin{align*}
&\QRep(\aa, p) = \{ \pi_p; \Pi_p \in \Rep(\aa_p)\},\\
&\Rep(\aa,p)= \{\pi_p \in \QRep(\aa, p); \pi_p \mbox{ is a
*-representation}\}, \\
&\FQRep(\aa, p) = \{ \pi_p ; \Pi_p \in \FRep(\aa_p) \}.
\end{align*}

\

{\bf Definition 2.7.} Let $p \in \Sigma (\aa)$. We say that $p$ is {\it
representable} if $\FRep(\aa, p) \equiv \{ \pi_p \in \FQRep(\aa, p); \pi_p
\text{ is a $*$-representation of } \aa \} \neq \emptyset$.

\ \\
It is natural to look for conditions for $p$ to be representable.
We shall consider this problem in Section 3.
\\
We define the notions of semifiniteness and weak semifiniteness of unbounded
C$^*$-seminorms, and study (quasi) $*$-representations induced by them.

\

{\bf Definition 2.8.} An unbounded $m^*$-seminorm $p$ on $\aa$ is said to be
{\it finite} if $\dd(p) = \N_p$; $p$ is said to be {\it semifinite} if $\N_p$
is $p$-dense in $\dd(p)$.
An unbounded C$^*$-seminorm $p$ on $\aa$ having Property (B) is said to be
{\it weakly semifinite} if
\begin{align*}
\QRep^\WB (\aa, p)
&\equiv \{ \pi \in \FQRep (\aa, p) ; \hh_{\pi_p} = \hh_{\Pi_p} \} \\
&\neq \emptyset.
\end{align*}
and an element $\pi_p$ of $\QRep^\WB (\aa, p)$ is said to be a {\it
well-behaved} quasi $*$-representation of $\aa$ in $\QRep(\aa, p)$.
A representable unbounded C$^*$-seminorm $p$ on $\aa$ having Property (B) is
said to be {\it weakly semifinite} if
\begin{align*}
\Rep^\WB (\aa, p)
&\equiv \QRep^\WB(\aa, p) \cap \Rep (\aa, p) \\
&\neq \emptyset.
\end{align*}
We remark that semifinite unbounded $m^*$- (or C$^*$-) seminorms automatically
satisfy Property (B).

\
Let $\pi$ be a (quasi) $*$-representation of $\aa$. We put
\begin{align*}
\aa^\pi_b = \{ x \in \aa; \overline{\pi(x)} \in \bb(\hh_\pi) \}, \\
\N_\pi \equiv \{ x \in \aa^\pi_b \cap R(\aa); ax \in \aa^\pi_b, \forall a \in
\aa \}.
\end{align*}

\

{\bf Definition 2.9.}
If $\pi(\aa) \dd(\pi)$ is total in $\hh_\pi$, then $\pi$ is said to be {\it
nondegenerate}.  If $\pi(\N_\pi) \dd(\pi)$ is total in $\hh_\pi$, then $\pi$
is said to be {\it strongly nondegenerate}.

\

{\bf Proposition 2.10.} Let $p$ be an unbounded C$^*$-seminorm on $\aa$ having
Property (B).
Then the following statements hold:

(1) $\QRep^\WB(\aa, p) \subset \{ \pi_p \in \QRep (\aa, p) ; \Pi_p \text{ is
nondegenerate} \}$, \\
\hspace*{12mm}
$\Rep^\WB(\aa, p) \subset \{ \pi_p \in \Rep (\aa, p) ; \Pi_p \text{ is
nondegenerate} \}$. \\
In particular, if $p$ is semifinite, then it is weakly semifinite and
\begin{align*}
& \QRep^\WB(\aa, p) = \{ \pi_p \in \QRep (\aa, p) ; \Pi_p \text{ is
nondegenerate} \}, \\
& \Rep^\WB(\aa, p) = \{ \pi_p \in \Rep (\aa, p) ; \Pi_p \text{ is
nondegenerate} \}.
\end{align*}
(2) Suppose $\pi_p \in \QRep^\WB (\aa, p)$.
Then

(i) $\pi_p(\N_p) \dd(\pi_p)$ is total in $\hh_{\pi_p}$, and so $\pi_p$ is
strongly nondegenerate,

(ii) $\| \overline{\pi_p(x)} \| = p(x)$ for each $x \in \dd(p)$,

(iii) $\pi_p(\aa)'_{\rm w} = \overline{\pi_p(\dd(p))}' \text{ and } \pi_p(\aa)
'_{\rm w} \dd(\pi_p) \subset \dd(\pi_p)$.\\
Conversely suppose $\pi_p \in \QRep(\aa, p)$ (resp. $\Rep(\aa,p)$) satisfies
conditions (i) and (ii) above. Then there exists an element $\pi_p^\WB$ of
$\QRep^\WB(\aa, p)$ (resp. $\Rep^\WB(\aa, p)$) which is a restriction of
$\pi_p$.\\[3mm]
\indent
{\bf Proof.} (1) Take an arbitrary $\pi_p \in \QRep^\WB(\aa, p)$.
Then since
\[
\dd(\pi_p) \subset \text{ linear span of } \Pi_p(\aa_p) \hh_{\Pi_p} \subset
\hh_{\Pi_p} = \hh_{\pi_p},
\]
it follows that $\Pi_p$ is nondegenerate.
Suppose $p$ is semifinite.
Let $\Pi_p \in \Rep(\aa_p)$ be nondegenerate.
Since $p$ is semifinite, it follows that $\{ \Pi_p(\tilde{x}); x \in \N_p \}$
is uniformly dense in the C$^*$-algebra $\Pi_p(\aa_p)$, which implies
$\hh_{\Pi_p} = \hh_{\pi_p}$.\\
(2) Let $\pi_p \in \QRep^\WB (\aa, p)$.
Since $\hh_{\pi_p} = \hh_{\Pi_p}$ and $\pi_p(x) = \Pi_p(\tilde{x}) \up
\dd(\pi_p)$ for each $x \in \dd(p)$, it follows that $\pi_p(\N_p) \dd(\pi_p)$
is total in $\Pi_p(\tilde{\N_p}) \hh_{\Pi_p}$ and $\N_p \subset \N_{\pi_p}$,
which implies by Lemma 2.4 (3) that the statement (i) holds.
Further, we have
\begin{equation}
\overline{\pp(x)} = \Pp(\tilde{x}), \hspace{5mm}x \in \dd(p),
\end{equation}
and hence
\[
\| \overline{\pp(x)} \| = \| \Pp(\tilde{x}) \| = p(x), \hspace{5mm} x \in
\dd(p).
\]
We next show the statement (iii).
Take an arbitrary $C \in \overline{\pp(\dd(p))}'$.
By (2.1) we have
\[
C \Pp(\tilde{x}) = C \overline{\pp(x)} = \overline{\pp(x)} C = \Pp(\tilde{x})
C
\]
for each $x \in \dd(p)$, which implies that $C \Pp(\tilde{x_1} \tilde{x_2})
\xi \in \dd(\pp)$ for each $x_1, x_2 \in \N_p$ and $\xi \in \hh_{\Pp}$ and
\begin{align*}
\pp(a) C \Pp(\tilde{x_1} \tilde{x_2}) \xi= \pp(a) \Pp(\tilde{x_1} \tilde{x_2})
C \xi
&= \Pp(({ax_1})^\sim) C \Pp(\tilde{x_2}) \xi \\
&= C \Pp(({ax_1})^\sim) \Pp(\tilde{x_2}) \xi \\
&= C \pp(a) \Pp(\tilde{x_1}\tilde{ x_2}) \xi
\end{align*}
for each $a \in \aa, x_1, x_2 \in \N_p$ and $\xi \in \hh_{\Pp}$.
Hence, $C \in \pp(\aa)'_{\rm w}$ and $C\dd(\pp) \subset \dd(\pp)$.
The converse inclusion $\pp(\aa)'_{\rm w} \subset \overline{\pp(\dd(p))}'$ is
trivial. Thus the statement (iii) holds.
Conversely suppose that $\pp \in \QRep(\aa, p)$ satisfies conditions (i) and
(ii). We put
\[
\Pp^\WB (\tilde{x}) = \overline{\pp(x)}, \hspace{5mm} x \in \dd(p).
\]
Then it follows from (ii) that
\[
\| \Pp^\WB (\tilde{x}) \| = \| \overline{\pp(x)} \| = p(x) = \| \tilde{x} \|_p
\]
for each $x \in \dd(p)$, and hence $\Pp^\WB$ can be extended to a faithful
$*$-representation of the C$^*$-algebra $\aa_p$ on the Hilbert space
$\hh_{\Pp^\WB}= \hh_{\pp}$. We denote it by the same symbol $\Pp^\WB$ and
denote by $\pp^\WB$ the quasi $*$-representation of $\aa$ induced by
$\Pp^\WB$. Then it follows from Lemma 2.4, (3) and statement (i) that
\begin{align*}
\hh_{\pp^\WB}
&= \text{ closed linear span of } \Pp^\WB (\tilde{\N_p}) \hh_{\Pp^\WB} \\
&=\text{ closed linear span of } \overline{\pp(\N_p)} \hh_{\pp} \\
&= \hh_{\pp} = \hh_{\Pp^\WB},
\end{align*}
so that $\pp^\WB \in \QRep^\WB (\aa, p)$.
Further, since $\Pp^\WB (\tilde{x}) = \overline{\pp(x)} = \Pp(\tilde{x}) \up
\hh_{\pp^\WB}$ for each $x \in \dd(p)$, it follows that $\pp^\WB$ is a
restriction of $\pp$.
Suppose $\pp \in \Rep(\aa, p)$.
Then, since $\pp^\WB$ is a restriction of $\pp$, it follows that $\pp^\WB$ is
a $*$-representation of $\aa$. This completes the proof.

\ \\
The set $\Sigma_B (\aa)$ of all unbounded C$^*$-seminorms on $\aa$ having
Property (B) is an ordered set with respect to the order relation $\subset$
defined by: $p \subset q$ if $\dd(p) \subset \dd(q)$ and $p(x) = q(x), \forall
x \in \dd(p)$.

\

{\bf Proposition 2.11.} Let $p$ and $q$ be in  $\Sigma_B(\aa)$. Suppose $p
\subset q$. Then, for any $\pp \in \QRep(\aa, p)$ there exists an element
$\pi_q$ of $\QRep(\aa, q)$ such that $\pp \subset \pi_q$.\\[3mm]
\indent
{\bf Proof.} Let $\aa_q$ be the C$^*$-algebra construted applying Lemma 2.2 to
$\dd(q)$.
 Then it follows from $p \subset q$ that for each $x \in \dd(p)$ we can define
\[
\Phi : \tilde{x} \in \widetilde{\dd(p)} \mapsto \tilde{x} \in
\widetilde{\dd(q)}.
\]
Then $\Phi$ is an isometric $*$-isomorphism of the dense subspace
$\widetilde{\dd(p)}$ of the C$^*$-algebra $\aa_p$ into the C$^*$-algebra
$\aa_q$, and so it can be extended to a $*$-isomorphism of the C$^*$-algebra
$\aa_p$ into the C$^*$-algebra $\aa_q$; we denote this extension by the same
symbol $\Phi$.
Take an arbitrary $\Pp \in \Rep(\aa_p)$. Since $\Pp \circ \Phi^{-1}$ is a
faithful $*$-representation of the C$^*$-algebra $\Phi(\aa_p)$ on $\hh_{\Pp}$
and every C$^*$-algebra is stable [14, Prop.2.10.2], it follows that $\Pp
\circ \Phi^{-1}$ can be extended to a $*$-representation $\Pi_q$ of the
C$^*$-algebra $\aa_q$ on $\hh_{\Pi_q}$, that is, $\hh_{\Pp}$ is a closed
subspace of $\hh_{\Pi_q}$ and $\Pi_q(\Phi(A)) \up \hh_{\Pp} = \Pp(A)$ for each
$A \in \aa_p$. Let $\pi_q$ denote the element of $\QRep(\aa, q)$ induced by
$\Pi_q$. Then we have
\begin{align*}
\pp(a) \Pp(\tilde{x_1} \tilde{x_2} ) \xi
&= \Pp((ax_1)^\sim) \Pp(\tilde{x_2}) \xi \\
&= \Pi_q(\Phi((ax_1)^\sim) ) \Pi_q (\Phi(\tilde{x_2})) \xi \\
&= \Pi_q((ax_1)^\sim \tilde{x_2}) \xi \\
&= \pi_q(a) \Pi_q(\tilde{x_1} \tilde{x_2}) \xi \\
&= \pi_q(a) \Pp(\tilde{x_1} \tilde{x_2}) \xi
\end{align*}
for each $a \in \aa$, $x_1, x_2 \in \N_p$ and $\xi \in \hh_{\Pp}$, and so $\pp
\subset \pi_q$. This completes the proof.
\section{Representability of unbounded C$^*$-seminorms}
Let $\aa$ be a partial $*$-algebra and $p$ an unbounded C$^*$-seminorm on
$\aa$. In this section we give some conditions under which the equality
$\Rep(\aa, p)= \QRep(\aa, p)$ holds.
The first case we consider is that of a semi-associative partial *-algebra
$\aa$.

\

{\bf Lemma 3.1.} Suppose that $\aa$ is a semi-associative partial $*$-algebra
$\aa$ and let $p \in \Sigma_B(\aa) $. Then $\Rep(\aa, p)= \QRep(\aa,
p)$.\\[3mm]
\indent
{\bf Proof.}
Since $\aa$ is semi-associative, it follows that
\[
y^* ((ab) x) = y^* (a(bx)) = (y^* a)(bx)
\]
for each $a \in L(b)$ and $x,y \in \N_p$, which implies
\begin{align*}
(\pp(ab) \Pp(\tilde{x_1} \tilde{x_2}) \xi | \Pp(\tilde{y_1} \tilde{y_2}) \eta)
&= (\Pp((y^*_1((ab)x_1))^\sim) \Pp(\tilde{x_2}) \xi | \Pp(\tilde{y_2}) \eta)
\\
&= (\Pp((y^*_1 a)^\sim (bx_1)^\sim) \Pp(\tilde{x_2}) \xi | \Pp(\tilde{y_2})
\eta) \\
&=(\pp(b) \Pp(\tilde{x_1} \tilde{x_2} ) \xi | \pp(a^*) \Pp(\tilde{y_1}
\tilde{y_2} ) \eta)\end{align*}
for each $x_1, x_2, y_1, y_2 \in \N_p$ and $\xi, \eta \in \hh_{\Pp}$. Hence
$\pp$ is a $*$-representation of $\aa$.

\

We next consider the case of (everywhere defined) C$^*$-seminorms.
Semi-associativity of $\aa$ is no more needed.

\

{\bf Lemma 3.2.} Let $\aa$ be a partial $*$-algebra. Suppose that $p$ is a
semifinite C$^*$-seminorm on $\aa$. Then $\Rep(\aa, p)= \QRep(\aa, p)$ and
every $\pp$ in $\Rep(\aa, p)$ is bounded.\\[3mm]
\indent
{\bf Proof.} Since $p$ is a C$^*$-seminorm on $\aa$, we have $\dd(p) = \aa$
and $\N_p = R(\aa)$. For any $a \in \aa$ we have $\pp(a) = \Pp(\tilde{a}) \up
\dd(\pp)$, and so $\pp(a)$ is bounded. Take arbitrary $a, b \in \aa$ s.t. $a
\in L(b)$. Then there exist sequences $\{ x_n \}$ and $\{ y_n \}$ in $R(\aa)$
such that $\{ x_n \}^\sim = \tilde{a}$ and $\{ y_n \}^\sim = \tilde{b}$, and
hence it follows from Lemma 2.2, (2) and Property (A) that
\begin{align*}
(\pp(ab) \Pp(\tilde{x_1}\tilde{x_2} ) \xi | \Pp(\tilde{y_1} \tilde{y_2} )
\eta) \\
&= (\Pp(\{ x_n y_n \}^\sim \tilde{x_1} ) \Pp(\tilde{x_2}) \xi |
\Pp(\tilde{y_1} \tilde{y_2} ) \eta) \\
&= (\Pp( \{ x_n \}^\sim) \Pp(\{ y_n \}^\sim ) \Pp(\tilde{x_1}\tilde{x_2} ) \xi
| \Pp(\tilde{y_1} \tilde{y_2} ) \eta) \\
&= (\pp(b)  \Pp(\tilde{x_1}\tilde{x_2} ) \xi | \pp(a^*) \Pp(\tilde{y_1}
\tilde{y_2} ) \eta)
\end{align*}
for each $x_1, x_2, y_1, y_2 \in \N_p$ and $\xi, \eta \in \hh_{\Pp}$. Hence,
$\pp$ is a $*$-representation of $\aa$.

\

{\bf Lemma 3.3.} Let $\aa$ be a partial $*$-algebra $\aa$ and $p \in
\Sigma_B(\aa)$. Assume that there exists a semifinite C$^*$-seminorm $\hat{p}$
on $\aa$ such that $p \subset \hat{p}$. Then $\Rep(\aa, p) = \QRep(\aa, p)$.
\\[3mm]
\indent
{\bf Proof.}
Take an arbitrary $\pp \in \QRep(\aa, p)$. By Proposition 2.11 and Lemma 3.2
there exists an element $\pi_{\hat{p}}$ of $\QRep(\aa, \hat{p})= \Rep(\aa,
\hat{p})$ such that $\pp \subset \pi_{\hat{p}}$, which implies that $\pp \in
\Rep(\aa, p)$.

\ \\
We consider now the special case of topological partial $*$-algebras. The
simplest situation is of course that of topological quasi $*$-algebras, where
we start from.

\

{\bf Lemma 3.4.} Suppose that $\aa$ is a topological quasi $*$-algebra over
$\aa_0$ and $p$ is an unbounded C$^*$-seminorm on $\aa$ having Property (B).
Then $\Rep(\aa, p) = \QRep(\aa, p)$. \\[3mm]
\indent
{\bf Proof.} Since every topological quasi $*$-algebra $\aa$ over $\aa_0$ is
semi-associative and $R(\aa) = \aa_0$, it follows from Lemma 3.1 that
$\Rep(\aa, p) = \QRep(\aa, p)$.

\ \\
Let $\aa[\tau]$ be a topological partial $*$-algebra and $p$ an unbounded
C$^*$-seminorm on $\aa$. For any $x \in \N_p$ we define a seminorm $p_x$ on
$\aa$ by
\[
p_x(a) = p(ax), \hspace{5mm} a \in \aa.
\]
We denote by $\tau_p$ the locally convex topology on $\aa$ defined by the
family $\{ p_x ; x \in \N_p \}$ of seminorms.
If $\tau_p \prec \tau$, then $p$ is said to be {\it locally continuous}.

\

{\bf Lemma 3.5.} Let $\aa[\tau]$ be a topological partial $*$-algebra
satisfying the following condition (C):

(C) For any $a \in \aa$, the linear map $L_a$ on $R(a)$ defined by $x \in R(a)
\mapsto ax \in \aa$ is continuous. \\
Suppose that $p$ is a locally continuous unbounded C$^*$-seminorm on $\aa$
having Property (B) and that $R(\aa) \cap \dd(p)$ is $\tau$-dense in $\aa$.
Then $\Rep(\aa, p) = \QRep(\aa, p)$.\\[3mm]
\indent
{\bf Proof.} Take arbitrary $a, b \in \aa$ s.t. $a \in L(b)$. Since $R(\aa)
\cap \dd(p)$ is $\tau$-dense in $\aa$, there exists a net $\{ y_{{}_\beta} \}$
in $R(\aa) \cap \dd(p)$ such that $\dis \tau-\lim_\beta y_{{}_\beta} = b$.
Further, since $\aa$ satisfies the condition (C), we have
$\tau$-$\dis\lim_\beta ay_{{}_\beta}= ab$, and since $p$ is locally
continuous, it follows that $\dis\lim_\beta p(y_{{}_\beta}x-bx)=0$ and
$\dis\lim_\beta p((ay_{{}_\beta})x-(ab)x)=0$ for each $x \in \N_p$. Hence we
have
\begin{align*}
(\pp(b) \Pp(\tilde{x_1} \tilde{x_2} ) \xi | \pp(a^*) \Pp(\tilde{y_1}
\tilde{y_2} ) \eta) \\
&= (\Pp((bx_1)^\sim \tilde{x_2}) \xi | \pp(a^*) \Pp(\tilde{y_1} \tilde{y_2} )
\eta) \\
&= \lim_\beta( \Pp((y_{{}_\beta} x_1)^\sim \tilde{x_2} ) \xi | \pp(a^*) \Pp
(\tilde{y_1} \tilde{y_2} ) \eta) \\
&= \lim_\beta(\Pp(\tilde{y_{{}_\beta}} \tilde{x_1}) \Pp(\tilde{x_2}) \xi |
\pp(a^*) \Pp(\tilde{y_1} \tilde{y_2} ) \eta) \\
&= \lim_\beta(\Pp((ay_{{}_\beta})^\sim \tilde{x_1} )  \Pp(\tilde{x_2}) \xi |
\Pp(\tilde{y_1} \tilde{y_2} ) \eta) \\
&= (\Pp(((ab) x_1)^\sim ) \Pp(\tilde{x_2}) \xi | \Pp(\tilde{y_1} \tilde{y_2} )
\eta) \\
&=(\pp(ab) \Pp(\tilde{x_1} \tilde{x_2} ) \xi | \Pp(\tilde{y_1} \tilde{y_2} )
\eta)
\end{align*}
for each $x_1, x_2, y_1, y_2 \in \N_p$ and $\xi, \eta \in \hh_{\Pp}$, which
implies that $\pp$ is a $*$-representation of $\aa$.
\section{Unbounded C$^*$-seminorms defined by $*$-represen-
tations}
In the previous sections we constructed $*$-representations of a partial
$*$-algebra $\aa$ from a representable unbounded C$^*$-seminorm on $\aa$
having Property (B).
Now, starting from a $*$-representation $\pi$ of $\aa$, we try to construct a
representable unbounded C$^*$-seminorm $r_\pi \in \Sigma_B(\aa)$. When this is
possible, it makes sense to investigate on the relation between $\pi$ and the
natural $*$-representation $\pi^N_{r_\pi}$ of $\aa$ induced by $r_\pi$
\\
Let $\pi$ be a $*$-representation of $\aa$ on a Hilbert space $\hh_\pi$. We
put, as above,
$$ \aa^\pi_b = \{ x \in \aa; \overline{\pi(x)} \in \bb(\hh_\pi) \}, $$
and
$$\pi_b(x) = \overline{\pi(x)}, \hspace{5mm} x \in \aa^\pi_b.$$
Then $\aa^\pi_b$ is a partial $*$-subalgebra of $\aa$ and $\pi_b$ is a bounded
$*$-representation of $\aa^\pi_b$ on $\hh_\pi$. An unbounded C$^*$-seminorm
$r^L_\pi$ on $\aa$ is defined by
\[
\dd(r^L_\pi) = \aa^\pi_b \text{ and }
r^L_\pi(x) = \| \pi_b(x) \|, \hspace{5mm} x \in \dd(r^L_\pi).
\]
But, $r^L_\pi$ does not necessarily have Property (B). For this reason, we
consider the family of all unbounded C$^*$-seminorms on $\aa$ having Property
(B) which are restrictions of $r^L_\pi$.
We denote this family by $\Sigma_B(\pi)$ and call it the {\it family of
unbounded C$^*$-seminorms induced by} $\pi$.

\

{\bf Definition 4.1.} If $\Sigma_B(\pi) \neq \{ 0 \}$, then $\pi$ is said to
have {\it Property (B)}.

\

Suppose that $\pi$ has Property (B) and $r_\pi \in \Sigma_B(\pi)$. We put
\[
\Pi(\tilde{x}) = \pi(x), \hspace{5mm} x \in \dd(r_\pi).
\]
Then since
\[
\| \Pi(\tilde{x}) \| = r_\pi(x) = \| \tilde{x} \|_{r_\pi}
\]
for each $x \in \dd(r_\pi)$, it follows that $\Pi$ can be extended to a
faithful $*$-representation
$\Pi^N_{r_\pi}$ of the C$^*$-algebra $\aa_{r_\pi} \equiv
\widehat{\dd(r_\pi)}/\sim$ on the Hilbert space $\hh_\pi$, and
$\Pi^N_{r_\pi}(\aa_{r_\pi}) = \overline{\pi(\dd(r_\pi))}^{\| \ \|}$.
We denote by $\pi^N_{r_\pi}$ the quasi $*$-representation of $\aa$ constructed
by $\Pi^N_{r_\pi}$. This is called the {\it natural representation of $\aa$
induced by $\pi$}.
Since $\hh_{\Pi^N_{r_\pi}} = \hh_\pi$, it follows that $\hh_{\pi^N_{r_\pi}}$
is a closed subspace of $\hh_\pi$.

\

{\bf Proposition 4.2.} Suppose that $\pi$ is a $*$-representation of $\aa$
having Property (B) and $r_\pi \in \Sigma_B(\pi)$. Then $r_\pi$ is
representable, $\pi^N_{r_\pi} \in \Rep(\aa, r_\pi)$ and $\hat{\pi} \up
\dd(\pi^N_{r_\pi})= \pi^N_{r_\pi}$.\\[3mm]
\indent
{\bf Proof.} Since
\begin{align*}
\dd(\pi^N_{r_\pi})
&= \text{ linear span of } \{ \Pi^N_{r_\pi} (\tilde{x_1} \tilde{x_2}) \xi ;
x_1, x_2 \in \N_{r_\pi}, \xi \in \hh_\pi \} \\
&= \text{ linear span of } \{ \overline{\pi(x_1 x_2) } \xi ; x_1, x_2 \in
\N_{r_\pi}, \xi \in \hh_\pi \},
\end{align*}
it follows that
\begin{align*}
(\pi(a)^* \eta | \Pi^N_{r_\pi}(\tilde{x_1} \tilde{x_2}) \xi )
&= (\pi(a)^* \eta | \overline{\pi(x_1 x_2)} \xi ) \\
&= (\overline{\pi((a x_1)^* ) } \eta | \overline{\pi(x_2)} \xi ) \\
&=(\eta | \overline{\pi((ax_1)x_2) } \xi ) \\
&=(\eta | \pi^N_{r_\pi} (a) \Pi^N_{r_\pi}(\tilde{x_1} \tilde{x_2}) \xi )
\end{align*}
for each $a \in \aa, \eta \in \dd(\pi(a)^*), x_1, x_2 \in \N_{r_\pi}$ and $\xi
\in \hh_\pi$, which implies that $\Pi^N_{r_\pi}(\tilde{x_1} \tilde{x_2})$
$\xi$ $\in \dd(\overline{\pi(a)})$ and $\overline{\pi(a)} \Pi^N_{r_\pi}(
\tilde{x_1} \tilde{x_2}) \xi = \pi^N_{r_\pi}(a) \Pi^N_{r_\pi}(\tilde{x_1}
\tilde{x_2}) \xi$.
Hence, $\dd(\pi^N_{r_\pi})$ $\subset$ $\dd(\hat{\pi})$ and $\hat{\pi} \up
\dd(\pi^N_{r_\pi})= \pi^N_{r_\pi}$, which implies since $\hat{\pi}$ is a
$*$-representa-tion of $\aa$ that $\pi^N_{r_\pi}$ is a $*$-representation of
$\aa$ and $r_\pi$ is representable. This completes the proof.

\ \\
We summarize in the following scheme the method of construction of
$\pi^N_{r_\pi}$ described above:

\vskip-12mm

\hspace*{25mm}\setlength{\unitlength}{0.7mm}
\begin{picture}(95,40)
\put(-30,0){\makebox(6,6)[l]{$\pi$}}
\put(25,0){\makebox(16,6)[l]{$\forall r_\pi \in \Sigma_B(\pi)$}}
\put(-40,-5){\makebox(6,6)[l]{$*$-representation}}
\put(-40,-10){\makebox(16,6)[l]{having Property (B)}}
\put(90,0){\makebox(10,6)[l]{$\aa_{r_\pi}$}}
\put(90,-5){\makebox(10,6)[l]{C$^*$-algebra}}
\put(4,2.5){\vector(1,0){20}}
\put(65,2.5){\vector(1,0){20}}
\put(-30,-18){\vector(0,1){0.5}}
\multiput(-30,-19)(0, -0.5){40}{\line(0,-1){0.4}}
\put(-30,-39){\vector(0,-1){0.5}}
\put(100,-10){\vector(0,-1){35}}
\put(105,-30){\makebox(10,6)[l]{$\Pi^N_{r_\pi}$}}
\put(-35, -55){\makebox(10,6)[l]{$\pi^N_{r_\pi}$}}
\put(90,-55){\makebox(10,6)[l]{$\overline{\pi_b(\dd(r_\pi))}^{ \| \ \|} $}}
\put(90,-65){\makebox(10,6)[l]{C$^*$-algebra on $\hh_\pi$}.}
\put(80,-50){\vector(-1,0){100}}
\end{picture}

\vskip5.5cm

{\bf Proposition 4.3.}
Suppose that $p$ is a representable weakly semifinite unbounded C$^*$-seminorm
on $\aa$ having Property (B). Then every $\pi_p$ of $\Rep^\WB(\aa, p)$ has
Property (B). Take an arbitrary $r_{\pi_p} \in \Sigma_B(\pi_p)$ which is an
extension of $p$. Then $\pp \subset \pi^N_{r_{\pp}}$ and $\hat{\pp}= \hat{\pi}
_{r^N_{\pp}}$.\\[3mm]
\indent
{\bf Proof.} Since $p \in \Sigma_B(\aa)$ and $p \subset r_{\pp} \subset
r^N_{\pp}$, it follows that $\pp$ has Property (B) and $\N_p \subset
\N_{r_{\pp}} \subset \aa^{\pp}_b$, which implies
\begin{align*}
\dd(\pp) &= \text{ linear span of } \{ \overline{\pp(x_1 x_2)} \xi ; x_1, x_2
\in \N_p, \xi \in \hh_{\pp} \} \\
&\subset  \text{ linear span of } \{ \overline{\pp(x_1 x_2)} \xi ; x_1, x_2
\in \N_{r_{\pp}}, \xi \in \hh_{\pp} \} \\
&= \dd(\pi^N_{\pp}) \subset \hh_{\pp}
\end{align*}
and $\pp= \pi^N_{r_{\pp}} \up \dd(\pp)$. On the other hand, it follows from
Proposition 4.2 that $\pi^N_{r_{\pp}} \subset \hat{\pp}$. Hence it follows
that $\hh_{\pp} = \hh_{\pi^N_{r_{\pp}}}, \pp \subset \pi^N_{r_{\pp}}$ and
$\hat{\pp} = \hat{\pi}{}^N_{r_{\pp}}$. This completes the proof.
\section{Examples}
In this section we give some examples of unbounded C$^*$-seminorms on partial
$*$-algebras having Property (B).

\

{\bf Example 5.1.} Let $S$ be a vector space of complex sequences containing
$l^\infty$. Suppose that $\{ x_n \}^* \equiv \{ \overline{x_n} \} \in S$ if
$\{ x_n \} \in S$ and $S$ is $l^\infty$-module. Then $S$ is a partial
$*$-algebra under the following partial multiplication and the involution: $\{
x_n \} \in L(\{ y_n \}) $ iff $\{ x_n y_n \} \in S$ and $\{ x_n \}^* = \{
\overline{x_n} \}$, and it has Property (A) and $R(S) \supset l^\infty$. We
define an unbounded C$^*$-norm on $S$ having Property (B) by
\[
\dd(r_\infty) = l^\infty \text{ and } r_\infty (\{ x_n \}) = \| \{ x_n \}
\|_\infty,
\hspace{5mm} \{ x_n \} \in \dd(r_\infty).
\]
For any $\{ x_n \} \in l^\infty$ we put
\[
\Pi_{r_\infty} (\{ x_n \}) \{ y_n \} = \{ x_n y_n \}, \hspace{5mm} \{ y_n \}
\in l^2.
\]
Then $\Pi_{r_\infty}$ is a faithful $*$-representation of the C$^*$-algebra
$l^\infty$ on the Hilbert space $l^2$ and since
\[
\N_{r_\infty} \supset \{ \{ x_n \} \in S \ ; x_n \neq 0 \text{ for only finite
numbers } n \},
\]
it follows that $\Pi_{r_\infty} (\N^2_{r_\infty}) \hh_{\Pi_{r_\infty}}$ is
total in $\hh_{\Pi_{r_\infty}}$. Hence $r_\infty$ is weakly semifinite.

\

{\bf Example 5.2.} Let $C(\rrr)$ be a $*$-algebra of all continuous
complex-valued functions on $\rrr$ equipped with the usual operations $f+g,
\lambda f, fg$ and the involution $f^*: f^*(t) = \overline{f(t)}, t \in \rrr$.
Let $\aa$ be a $*$-vector subspace of $C(\rrr)$ containing $C_b(\rrr) \equiv
\{ f \in C(\rrr) ; f \text{ is bounded} \}$ and
suppose that $\aa$ is $C_b(\rrr)$-module. Concrete examples of partial
$*$-algebras of this kind are, for instance:

(1) $\aa = C(\rrr)$.

(2) $ \aa = C(\rrr) \cap L^p(\rrr), \hspace{5mm} (1 \leq p < \infty)$.

(3) $ \aa = \aa_n \equiv \{ f \in C(\rrr); \dis \sup_{t \in \rrr} \dfrac{|f(t)
|}{(1+t^2)^n} < \infty \}, \hspace{5mm} n \in \nnn$.\\

Then $\aa$ is a partial $*$-algebra having Property (A) under the following
partial multiplication: $f \in L(g)$ iff $fg \in \aa$. We here define an
unbounded C$^*$-norm on $\aa$ by
\[
\dd(r_\infty) = C_b(\rrr) \text{ and } r_\infty(f) = \| f \|_\infty \equiv
\sup_{t \in \rrr} | f(t) |, \hspace{5mm} f \in \dd(r_\infty).
\]
Then, since $\dd(r_\infty) \subset R(\aa)$, it follows that $r_\infty$ has
Property (B).
Further, a faithful $*$-representation $\Pi_{r_\infty}$ of the C$^*$-algebra
$C_b(\rrr)$ on $L^2(\rrr)$ is defined by
\[
\Pi_{r_\infty}(f) g= fg, \hspace{5mm} f \in C_b(\rrr), g \in L^2(\rrr).
\]
Since $\N_{r_\infty} \supset C_c(\rrr) \equiv \{ f \in C(\rrr) ; {\rm supp} f
\text{ is compact }\}$, it follows that $\Pi_{r_\infty} (\N^2_{r_\infty})
L^2(\rrr)$ is total in $L^2(\rrr)$, which means that $r_\infty$ is weakly
semifinite.
Similarly we have the following

\

{\bf Example 5.3.} (1) Let $C^\infty(\rrr) \cap L^p(\rrr)$ $(1 \leq p <
\infty)$, where $C^\infty(\rrr)$ is a $*$-algebra of all infinitely
differentiable complex-functions on $\rrr$. We put
\[
\dd(r_\infty) = C^\infty_b(\rrr) \text{ and }
r_\infty(f) = \sup_{t \in \rrr} | f(t)|, \hspace{5mm} f \in \dd(r_\infty).
\]
Then $r_\infty$ is a weakly semifinite unbounded C$^*$-norm on $C^\infty(\rrr)
\cap L^p(\rrr)$ having Property (B). \\
(2) Let $n \in \nnn$ and $\aa_n = \{ f \in C^\infty(\rrr); \dis \sup_{t \in
\rrr} \dfrac{|f(t)|}{(1+t^2)^n} < \infty \}$. We put
\[
\dd(r_\infty) = C^\infty_b(\rrr) \text{ and } r_\infty(f) = \dis \sup_{t \in
\rrr} | f(t) | , \hspace{5mm} f \in \dd(r_\infty).
\]
Then $\aa_n$ is a partial $*$-algebra having Property (A) and $r_\infty$ is a
weakly semifinite unbounded C$^*$-norm on $\aa_n$ having Property (B).

\

{\bf Example 5.4.} Let $\dd$ be a dense subspace of a Hilbert space $\hh$ and
$C(\hh)$ the C$^*$-algebra of all compact operators on $\hh$.
Suppose that the maximal partial O$^*$-algebra $\ll^\dagger (\dd, \hh)$ is
self-adjoint.
Then $\ll^\dagger (\dd, \hh)$ has Property (A) and $R^{\rm w} (\ll^\dagger
(\dd, \hh)) = \{ X \up \dd; \overline{X} \in \bb(\hh) \text{ and }
\overline{X} \hh \subset \dd \}$. We now define an unbounded C$^*$-norm $r_u$
on the maximal O$^*$-algebra $\ll^\dagger (\dd, \hh)$ by
\[
\dd(r_u) = C(\hh) \up \dd \text{ and } r_u (X) = \| \overline{X} \|,
\hspace{5mm} X \in \dd(r_u).
\]
Since $F(\dd, \hh) \equiv \text{ linear span of } \{ \xi \otimes \overline{y}
; \xi \in \dd, y \in \hh \} \subset R^{\rm w} (\ll^\dagger (\dd, \hh))$, where
$(x \otimes \overline{y}) z= ( z | y )x$ for $x, y, z \in \hh$ and $F(\dd,
\hh)$ is uniformly dense in $C(\hh)$, it follows that $r_u$ has Property (B).
Further, since $\N_{r_\infty} \supset F(\dd, \hh)$, it follows that $r_u$ is
semifinite.

\

{\bf Example 5.5.} Let $\mm_0$ be an O$^*$-algebra on the Schwartz space
$\ss(\rrr)$ and $N= \dis \sum^\infty_{n=0} (n+1) f_n \otimes \overline{f_n} $
the number operator, where $\{ f_n \} \subset \ss(\rrr)$ is an orthonormal
basis in $L^2(\rrr)$ consisting the normalized Hermite functions. Let $\mm$ be
a partial O$^*$-algebra on $\ss(\rrr)$ containing $\mm_0$ and $\{ f_n \otimes
\overline{f_m} ; n, m \in \nnn \cup \{ 0 \} \}$. Since $\mm$ is self-adjoint,
it follows that $\mm$ has Property (A). We define an unbounded C$^*$-norm on
$\mm$ by
\[
\begin{cases}
\dd(r_u) = \text{ linear span of } \{ A f_n \otimes \overline{B f_m} ; A, B
\in \mm, n, m \in \nnn \cup \{ 0 \} \} \\
r_u (X) = \| \overline{X} \|, \hspace{5mm} X \in \dd(r_u).
\end{cases}
\]
Then, the linear span of $\{ f_n \otimes \overline{B f_m } ; B \in \mm, n, m
\in \nnn \cup \{ 0 \} \}$ is contained in $\N_{r_u}$ and it is uniformly dense
in $\dd(r_u)$.
Hence $r_u$ has Property (B) and it is semifinite.

\

{\bf Example 5.6.} Let $(\aa, \aa_0)$ be a proper CQ*-algebra
\cite{btcq1,btcq2}, i.e. a topological quasi *-algebra  $(\aa [\tau],
\aa_0)$ such that:

a)$\aa [\tau]$ is a Banach space under the norm $\|\,\|$;\\

b) the involution * of $\aa$ is isometric, i.e. $\|X\|=\|X^*\|, \;\forall X
\in \aa$;\\

c) $ \|X\|_0= \max \{ \|X\|_R, \|X^*\|_R  \}$ where $\|X\|_R =\sup\{\|AX\|;
\|A\|\leq 1$ \}. \\
Of course the C*-norm on $\aa_0$ can be viewed as an unbounded C*-norm $r$
with domain $\dd (r) =\aa_0$. Since $R(\aa)=\aa_0$, it is obvious that $r$
satisfies the property (B). In order to apply the results of Section 2, we
have to consider the set
$$ \N_r= \{X \in \aa_0: AX \in \aa_0, \; \forall A \in \aa\}. $$
Also in this simple situation, $\N_r$ might be trivial.
Let us sketch a concrete case, where this does not happen.\\
Let $S$ be an unbounded selfadjoint operator in Hilbert space $\hh $,
$S\geq 1$. The norm
$$\|X\|_S := \|S^{-1}XS^{-1}\|, \quad X \in \mathcal{B}(\hh)$$
defines a topology stricly weaker than the one defined by the C*-norm of
$\mathcal{B}(\hh)$.\\
Let $$C(S)= \{X \in \mathcal{B}(\hh): XS^{-1}=S^{-1}X\}.$$
$C(S)$ is a C*-algebra under the norm of $\mathcal{B}(\hh)$ and its
$\|\,\|_S$-completion $\widehat{C}(S)$ is a CQ*-algebra on $C(S)$
\cite[Proposition 2.6]{btcq2}.\\
Now define
$$\dd (r)=C(S) \hspace{3mm} \mbox{and } r(X)=\|X\|, \;\; X \in C(S),$$
then $r$ is an unbounded C*-norm on $\widehat{C}(S)$ satisfying the
property (B).
It is easy to check that
$$\N_r \supset  \{X \in C(S): X \mbox{ is of finite rank}\}.$$
So, for instance if $S$ has the spectral decomposition
$$ S=\sum_1^{\infty} \lambda_n P_n $$
where the $P_n$'s are finite rank projections, then $\N_r$ is non trivial
and the construction of Section 2 applies. We remark that $r$ need not be
semifinite, but, by Lemma 3.4, any element of $\QRep (\aa,p)$ is indeed a
*-representation.

\vspace{6mm}
{\noindent}{\bf Acknowledgment} Two of us (F.B and C.T) wish to acknowledge
the warm hospitatily of the Department of Applied Mathematics of the Fukuoka
University, where a part of this paper was performed.


\begin{thebibliography}{99}
\bibitem{alcyng}{\sc J. Alcantara and J. Yngvason}, Algebraic quantum field
theory and non commutative moment problem II, {\em Ann. Inst. Henri
Poincar{\'e}}, {\bf 48} (1988), 161-173.
%
\bibitem{abt}{\sc J.-P. Antoine, F.Bagarello and C.Trapani}, Topological
partial *-algebras: Basic properties and examples, {\em Rev. Math. Phys.} {\bf
11} (1999) 267 - 302
%
\bibitem{ak} {\sc J-P. Antoine and W. Karwowski}, Partial *-algebras of closed
          linear operators in Hilbert space,
      {\em   Publ. RIMS, Kyoto Univ.} {\bf 21}  (1985) 205-236;
%
\bibitem{ait1} {\sc J-P. Antoine, A. Inoue and C. Trapani}, Partial *-algebras
of
       closable operators. I. The basic theory and the abelian case,
     {\em  Publ. RIMS, Kyoto Univ.} {\bf 26} (1990) 359-395
%
\bibitem{ait2} {\sc J-P. Antoine, A. Inoue and C. Trapani}, Partial *-algebras
of
       closable operators.
    II. States and representations of partial *-algebras,
     {\em  Publ. RIMS, Kyoto Univ.} {\bf 27} (1991) 399-430
%
\bibitem{aitrev} {\sc J-P. Antoine, A. Inoue and C. Trapani}, Partial
*-algebras          of  closable operators: A review,
        {\em Reviews Math. Phys.} {\bf 8} (1996) 1-42
%
\bibitem{btcq1}{\sc F.Bagarello and C.Trapani}, States and representations
of $CQ^\ast$
-algebras, {\em Ann. Inst. H. Poincar\'e} {\bf 61}, (1994)
 103-133
%
\bibitem{btcq2}{\sc F.Bagarello and C.Trapani}, CQ*-algebras: structure
properties,{\em  Publ. RIMS, Kyoto Univ.} {\bf 32}, (1996) 85-116
%
\bibitem{bio2}{\sc S. J. Bhatt, A. Inoue and H. Ogi},{ On C*-spectral
algebras}, {\em Suppl. Rend. Circ. Mat.Palermo}, {\bf 56} (1998).
%
\bibitem{bio1}{\sc S. J. Bhatt, A. Inoue and H. Ogi},{ Admissibility of
weights on non-normed *-algebras}, {\em Trans. Amer. Math. Soc.}, {\bf 351}
(1999), 4629-4656.
%
\bibitem{bio3}{\sc S. J. Bhatt, A. Inoue and H. Ogi}, Unbounded C*-seminorms
and unbounded C*-spectral algebras, {\em preprint (Fukuoka Univ.} 1998).
%
\bibitem{bio4}{\sc S. J. Bhatt, A. Inoue and H. Ogi}, C$^*$-spectrality and
spectral invariance in locally convex $*$-algebras, {\em preprint (Fukuoka
Univ.} 1999).
%
\bibitem{vj}{\sc F. F. Bonsall and J. Duncan},
Complete Normed Algebras, {\em Springer-Verlag, Berlin, Heidelberg, New York},
1973.
%
\bibitem{dixmier}{\sc J. Dixmier},
C$^*$-Algebras, {\em North-Holland Publ. Comp.}, Amsterdam, 1977.
%
\bibitem{dubois}{\sc M. Dubois-Violette}, A generalization of the classical
moment problem on $*$-algebras with applications to relativistic quantum
theory I, {\em Commun. Math. Phys.}, {\bf 43} (1975) 225-254: II, {\em Commun.
Math. Phys.}, {\bf 54} (1977), 151-172.
%
\bibitem{lass1} {\sc G. Lassner}, Topological algebras and
        their applications in Quantum Statistics, {\em Wiss. Z. KMU-Leipzig,
     Math.-Naturwiss. R.} {\bf 30} (1981) 572-595
%
\bibitem{lass2} {\sc G. Lassner}, Algebras of unbounded operators
        and quantum dynamics, {\em Physica} {\bf 124~A} (1984) 471-480
%
\bibitem{yng}{\sc J. Yngvason}, Algebraic quantum field theory and non
commutative moment problem I, {\em Ann. Inst. Henri Poincar{\'e}}, {\bf 48}
(1988), 147-159.
%
\bibitem{yood}{\sc B. Yood}, C*-seminorms, {\em Studia Math.}, {\bf 118}
(1996), 19-26.





\end{thebibliography}
\end{document}